\documentclass[pmlr]{jmlr}

\usepackage{enumitem} 
\usepackage{stmaryrd} 
\usepackage{tikz}
\usetikzlibrary{arrows.meta, automata, positioning}

\tikzstyle{tr}=[-{>[length=3pt]}, shorten >=2pt]
\tikzset{every initial by arrow/.style={tr}}

\newcommand*{\lstar}{\textsf{L*}}
\newcommand*{\olstar}{\textsf{OL*}}
\newcommand*{\ttt}{\textsf{TTT}}

\newcommand*{\oneact}[1]{\mathbf{1}_{#1}\!\cdot\!}
\newcommand*{\one}[1]{\mathbf{1}_{#1}}
\DeclareMathOperator{\last}{\textsf{\upshape last}}
\newcommand*{\sem}[1]{\llbracket #1 \rrbracket}
\newcommand{\emptyword}{\varepsilon}

\jmlrworkshop{LearnAut 2024}

\title[Output-decomposed Learning of Mealy Machines]{Output-decomposed Learning of Mealy Machines}

\author{\Name{Rick Koenders}\Email{rick.koenders@ou.nl}\and
 \Name{Joshua Moerman}\Email{joshua.moerman@ou.nl}\\
 \addr Open Universiteit, the Netherlands}

\begin{document}
\maketitle

\begin{abstract}
We present an active automata learning algorithm which learns a decomposition of a finite state machine, based on projecting onto individual outputs.
This is dual to a recent compositional learning algorithm by~\citet{LabbafGHM23}.
When projecting the outputs to a smaller set, the model itself is reduced in size.
By having several such projections, we do not lose any information and the full system can be reconstructed.
Depending on the structure of the system this reduces the number of queries drastically, as shown by a preliminary evaluation of the algorithm.
\end{abstract}
\begin{keywords}
Active Automata Learning, Model Learning, Compositionality, Finite State Machines
\end{keywords}

\section{Introduction}
\label{sec:intro}

Model learning is an automated technique to construct a finite state machine for a closed-box system.
Using only observations based on input and output behaviour, model learning algorithms are successfully applied to systems, e.g., in order to find security flaws in DTLS~\citep{Fiterau-Brostean23}, or to understand legacy systems at ASML~\citep{SanchezGS19}.
Many more applications of model learning are listed by~\citet{Vaandrager17}.

In these applications, systems are assumed to behave like finite state machines, and algorithms such as \lstar{} and \ttt{} are very capable of inferring the state machine from observed inputs and outputs.
However, often times, systems are not \emph{just} a finite state machine; they are engineered in a structured way, by re-using common components or composing separate modules into a bigger system.
This additional structure is not used in these learning algorithms.
Only recently new learning algorithms are described which incorporate some sort of \emph{compositionality} into model learning, see for instance the papers by~\citet{LabbafGHM23, FrohmeS21, Moerman18}; and~\citet{NeeleS23}.

In this paper we take a dual approach to that of~\citet{LabbafGHM23}.
Instead of decomposing the set of inputs into independent smaller sets, we decompose the output.
This is similar to the approach of~\citet{Moerman18}, but with the advantage that our new approach does not assume a priori knowledge of the decomposition and is always applicable.

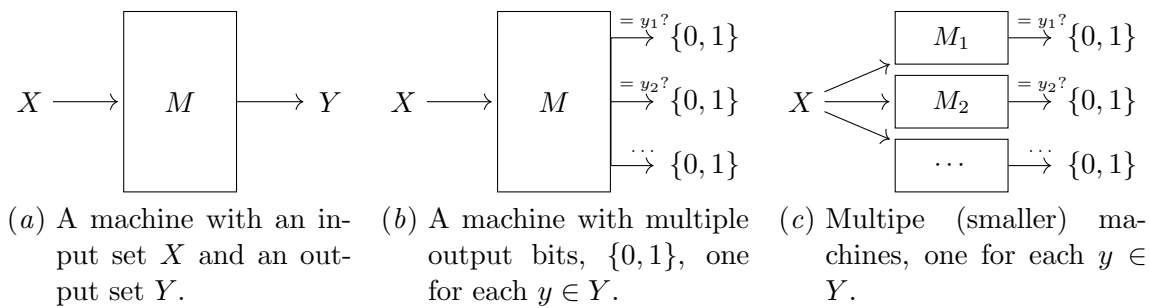
\begin{figure}[t]
\floatconts{fig:main-idea}
{\caption{The main idea of output decomposition.}}
{%
\subfigure[A machine with an input set $X$ and an output set $Y$.]{\label{fig:main-idea-a}%
  \begin{tikzpicture}
    \node (box) at (0, 0) [shape=rectangle, draw, minimum width=1.5cm, minimum height=2.4cm] {$M$};
    \node (l) at (-2, 0) {$X$};
    \node (r) at (2, 0) {$Y$};
    \draw[tr] (l) -- (box);
    \draw[tr] (box) -- (r);
  \end{tikzpicture}%
  }%
\quad
\subfigure[A machine with multiple output bits, $\{0, 1\}$, one for each $y \in Y$.]{\label{fig:main-idea-b}%
  \begin{tikzpicture}
    \node (box) at (0, 0) [shape=rectangle, draw, minimum width=1.5cm, minimum height=2.4cm] {$M$};
    \node (l) at (-2, 0) {$X$};
    \node (r1) at (2, .85) {$\{0, 1\}$};
    \node (r2) at (2, 0) {$\{0, 1\}$};
    \node (r3) at (2, -0.85) {$\{0, 1\}$};
    \draw[tr] (l) -- (box);
    \draw[tr] (box.east) |- node [pos=0.85, above] {\tiny $= y_1$?} (r1);
    \draw[tr] (box.east) |- node [pos=0.85, above] {\tiny $= y_2$?} (r2);
    \draw[tr] (box.east) |- node [pos=0.85, above] {\tiny $\cdots$} (r3);
  \end{tikzpicture}%
  }%
\quad
\subfigure[Multipe (smaller) machines, one for each $y \in Y$.]{\label{fig:main-idea-c}%
  \begin{tikzpicture}
    \node (box1) at (0, .85) [shape=rectangle, draw, minimum width=1.5cm, minimum height=.7cm] {\small $M_1$};
    \node (box2) at (0, 0) [shape=rectangle, draw, minimum width=1.5cm, minimum height=.7cm] {\small $M_2$};
    \node (box3) at (0, -.85) [shape=rectangle, draw, minimum width=1.5cm, minimum height=.7cm] {\small $\cdots$};
    \node (l) at (-2, 0) {$X$};
    \node (r1) at (2, .85) {$\{0, 1\}$};
    \node (r2) at (2, 0) {$\{0, 1\}$};
    \node (r3) at (2, -0.85) {$\{0, 1\}$};
    \draw[tr] (l) -- (box1);
    \draw[tr] (l) -- (box2);
    \draw[tr] (l) -- (box3);
    \draw[tr] (box1.east) |- node [pos=0.85, above] {\tiny $= y_1$?} (r1);
    \draw[tr] (box2.east) |- node [pos=0.85, above] {\tiny $= y_2$?} (r2);
    \draw[tr] (box3.east) |- node [pos=0.85, above] {\tiny $\cdots$} (r3);
  \end{tikzpicture}%
  }\vspace{-6pt}
}
\end{figure}

The main idea is simple: Take a Mealy machine with inputs from a set $X$ and outputs in a set $Y$ (\figureref{fig:main-idea-a}).
Then for each output $y \in Y$, we can imagine a separate output wire which indicates whether the current output equals $y$ or not (\figureref{fig:main-idea-b}).
These output wires have the property that exactly one output wire is active.
Observing each wire individually, we may learn a model of the state machine associated to that output only.
We call such a machine where we focus on only one output a \emph{projection}.
These models may be smaller than the whole system (\figureref{fig:main-idea-c}).

An example of decomposing the system this way can be seen in~\figureref{fig:example}. The automaton in~\figureref{fig:automaton} has a cyclic output for the input $a$, and the input $b$ reverses the cycle. This automaton can be decomposed into three smaller automata, one for each of the three outputs. The three projections all have 3 states, while the original automaton has 6 states. The projection on the output $x$ can be seen in~\figureref{fig:projection}. The other two projections (onto $y$ and $z$) have a similar, but different shape.
One might worry that learning three separate automata with 3 states would require more work, compared to learning a single 6-state automaton. Luckily, the three automata are not independent and the learning algorithm can re-use certain observations.
Our learning algorithm \olstar{} which learns the three 3-state automata requires fewer queries than learning the 6-state automaton with \lstar{}.

\newsavebox{\examplebox}
\savebox{\examplebox}{\begin{tikzpicture}[node distance = 2.6 cm, on grid, auto]
        \node (q0) [state, initial, initial text = {}] {$q_0$};
        \node (q1) [state, below of=q0] {$q_1$};
        \node (q2) [state, below of=q1] {$q_2$};
        \node (q3) [state, right of=q0] {$q_3$};
        \node (q4) [state, below of=q3] {$q_4$};
        \node (q5) [state, below of=q4] {$q_5$};
        \path [tr]
            (q0) edge node {$a/y$} (q1)
            (q1) edge node {$a/z$} (q2)
            (q2) edge [bend left, left] node {$a/x$} (q0)
            (q5) edge node {$a/y$} (q4)
            (q4) edge node {$a/x$} (q3)
            (q3) edge [bend left, right] node {$a/z$} (q5)
            (q0) edge [bend left] node {$b/x$} (q3)
            (q3) edge [bend left] node {$b/x$} (q0)
            (q1) edge [bend left] node {$b/y$} (q4)
            (q4) edge [bend left] node {$b/y$} (q1)
            (q2) edge [bend left] node {$b/z$} (q5)
            (q5) edge [bend left] node {$b/z$} (q2);
      \end{tikzpicture}}
\begin{figure}[t]
\floatconts
  {fig:example}
  {\caption{Example Mealy machine where its $x$-projection is smaller.}}
  {%
    \subfigure[Mealy machine with three outputs: $Y = \{x, y, z\}$.][t]{\label{fig:automaton}%
      \usebox{\examplebox}}
    \qquad
    \subfigure[Projection onto $x$ (minimised).][t]{\label{fig:projection}%
      \raisebox{\dimexpr.5\ht\examplebox-.5\height}{
      \begin{tikzpicture}[node distance = 2.6 cm, on grid, auto]
        \node (q0) [state, initial, initial text = {}] {$q^x_0$};
        \node (q1) [state, below of=q0] {$q^x_1$};
        \node (q2) [state, below of=q1] {$q^x_2$};
        \path [tr]
            (q0) edge node {$a/0$} (q1)
            (q1) edge node {$a/0$, $b/0$} (q2)
            (q2) edge [bend left=60, left] node {$a/1$} (q0)
            (q0) edge [loop right] node {$b/1$} (q0)
            (q2) edge [bend left, left] node {$b/0$} (q1);
    \end{tikzpicture}}}\vspace{-6pt}
  }
\end{figure}
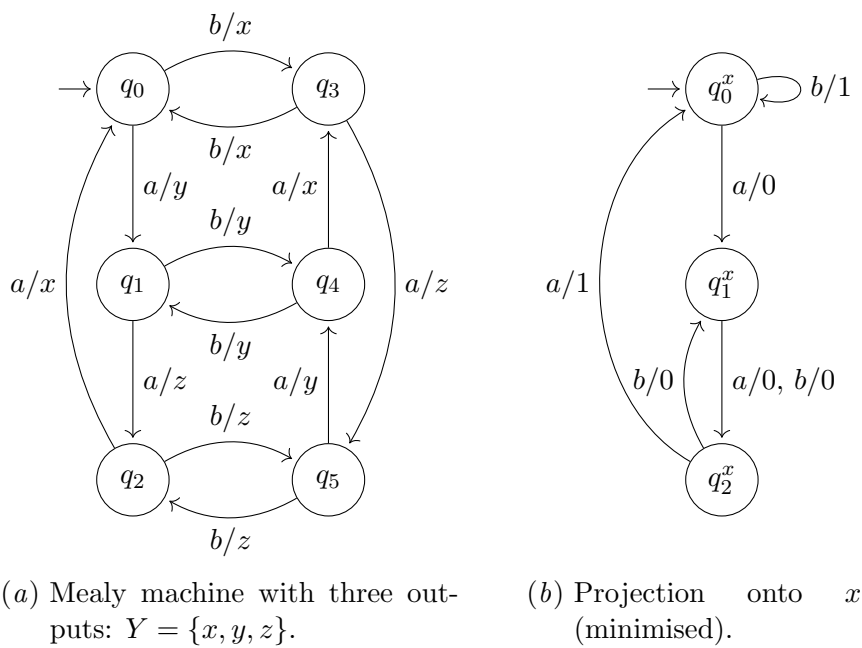

In practice, it varies how many states the projections have compared to the original automaton. In the worst case, one of the projections requires the same number of states as the original automaton. In this case, the benefits of our approach do not apply, and the algorithm may perform worse than \lstar{}. For automata where the projections are smaller than the full automaton, though, our approach can be much better than \lstar{}.

Our algorithm \olstar{} is based on the \lstar{} algorithm by~\citet{Angluin1987} and is still work-in-progress.
We expect that the improvements we see are transferable to more efficient algorithms such as \ttt{} by~\citet{IsbernerHS14}.

\section{Preliminaries}
\label{sec:prelim}

We recall the common definitions and notations on sets and words (or sequences).
The set of all words with symbols in $A$ is denoted by $A^*$, which includes the empty word $\emptyword{} \in A^*$.
The set of non-empty words is $A^+$.
The last symbol of a word $w \in A^+$ is denoted by $\last(w) \in A$.
Concatenation of two words $u, v$ is written as $u \cdot v$ or as $uv$.
This generalises to sets of words: $U \cdot V = \{uv \mid u \in U, v \in V\}$ for $U, V \subseteq A^*$.
Given a function $f \colon A \to B$, we get an induced function $f^* \colon A^* \to B^*$, where $f$ acts on each symbol.

\subsection{Machines}

We focus on deterministic finite state machines.

\begin{definition}[Mealy machine]\label{def:mealy}
    A \emph{Mealy machine} is a tuple $M = (Q, X, Y, q_0, \delta, \lambda)$ where
    $Q$ is a finite set of states;
    $X$ is a finite set of inputs;
    $Y$ is a finite set of outputs;
    $q_0 \in Q$ is the initial state;
    $\delta \colon Q \times X \to Q$ is the transition function; and
    $\lambda \colon Q \times X \to Y$ is the output function.
\end{definition}
In figures we depict transitions as $q \xrightarrow{a/y} q'$ whenever $\delta(q, a) = q'$ and $\lambda(q, a) = y$.
We extend $\delta$ and $\lambda$ to words in $I^*$ in the usual inductive way:
\begin{align*}
\delta^* \colon Q \times X^* &\to Q           & \lambda^* \colon Q \times X^* &\to Y^* \\
\delta^*(q, \emptyword) &= q                  & \lambda^*(q, \emptyword) &= \emptyword \\
\delta^*(q, aw) &= \delta^*(\delta(q, a), w)  & \lambda^*(q, aw) &= \lambda(q, a)\cdot\lambda^*(\delta(q, a), w)
\end{align*}
When convenient, we write $\delta$ and $\lambda$ instead of $\delta^*$ and $\lambda^*$.

\begin{definition}[Semantics of a Mealy machine]
    We define the \emph{semantics} of a Mealy machine $M = (Q, X, Y, q_0, \delta, \lambda)$ as the function $\sem{M} \colon X^* \to Y^*$, assigning to each input word its output:
    \[ \sem{M}(w) = \lambda^*(q_0, w) \]
\end{definition}

\subsection{Projections}

In general, we can compose the output function $\lambda$ of a Mealy machine with any function $f \colon Y \to Z$ to obtain a Mealy machine with outputs in $Z$ instead of $Y$.
Formally, given a Mealy machine $M = (Q, X, Y, q_0, \delta, \lambda)$ and a function $f \colon Y \to Z$, the \emph{composition} is $M^f = (Q, X, Z, q_0, \delta, f \circ \lambda)$.

\begin{lemma}
For a Mealy machine $M$ with output set $Y$ and function $f \colon Y \to Z$, we have:
\[ \sem{M^f} = f^* \circ \sem{M} \]
\end{lemma}

For our purpose of projecting onto a single output $y \in Y$, we use the function $\one{y}$ which is defined by $\one{y}(x) = 1$ if $x = y$ and $\one{y}(x) = 0$ otherwise.

\begin{definition}[Projection of Mealy machine]
Given a Mealy machine $M$ and an output $y \in Y$, its \emph{projection onto $y$} is defined by $M^{\one{y}}$ and will simply be denoted by $M^y$.
Its set of outputs is $\{0, 1\}$.
\end{definition}

\begin{remark}
Note that $M^f$ may have pairs of states which are behaviourally equivalent, while not being behaviourally equivalent in $M$.
This fact provides an opportunity to reduce the size of a system we are trying to learn.
\end{remark}

When composing with a function $f$, we might lose information, as certain outputs will not be distinguishable anymore.
With $\one{y}$ we almost definitely lose information because all outputs besides $y$ will be indistinguishable.
To cope with this, we consider the set of all projections $\{\one{y}\}_{y \in Y}$ and this ensures we can reconstruct $M$ from its projections.
In order to reconstruct the Mealy machine from its projections, we introduce the following type of composition, which takes the product of all the state spaces.

\begin{definition}[(Re-)Composition]\label{def:recomposition}
Let $I$ be an index set and $M_i = (Q_i, X, Y_i, q_{0,i}, \delta_i, \lambda_i)$ be Mealy machines for each $i \in I$ on a common input $X$.
Let $Y$ be any set and $f_i \colon Y \to Y_i$ be functions where $Y_i$ are the output sets of $M_i$.
We define the \emph{composition $\Pi_{i \in I} (M_i, f_i)$ of $M_i$ along $\langle f_i \rangle_{i \in I}$} to be a Mealy machine $\Pi_{i \in I} (M_i, f_i) = (\Pi_{i \in I} Q_i, X, Y, \langle q_{0,i} \rangle_{i \in I}, \delta', \lambda')$, where $\delta'$ and $\lambda'$ are given by the following rule:
\begin{center}
\vspace{-6pt}
\begin{tikzpicture}
\node at (0, 0) (top) {$ f_i (y) = y_i \quad \land \quad q_i \xrightarrow{a / y_i} q_i' \qquad \forall i \in I$};
\node at (0, -0.8) {$\langle q_j \rangle_{j \in I} \xrightarrow{a / y} \langle q_j' \rangle_{j \in I}$};
\draw (top.south west) -- (top.south east);
\end{tikzpicture}
\vspace{-6pt}
\end{center}
\end{definition}

Note that the rule for the transition structure is not always well-defined.
If the machines $M_i$ are constructed from $M^{f_i}$, we can ensure that there is at least one transition for each state and input.
Additionally, if the functions $f_i$ are jointly injective, then there is at most one transition for each state and input.
So under these restrictions, the composed Mealy machine is completely specified and deterministic.

\begin{definition}
A set of functions $\{f_i \colon Y \to Z_i\}_{i \in I}$ on a common domain is \emph{jointly injective} if $\langle f_i \rangle_{i \in I} \colon Y \to \Pi_{i \in I} Z_i$ is injective.
\end{definition}

\begin{lemma}[Reconstruction of $M$ from its projections]\label{lemma:reconstruction}
For a Mealy machine $M$, if the set $\{f_i\}_{i \in I}$ is jointly injective, then the Mealy machine $\Pi_{i \in I} (M^{f_i}, f_i)$ is well-defined and
\[ \sem{ \Pi_{i \in I} (M^{f_i}, f_i) } = \sem{M} \]
\end{lemma}

\begin{lemma}
The set $\{\one{y} \colon Y \to \{0, 1\}\}_{y \in Y}$ is jointly injective.
\end{lemma}

\subsection{\lstar{}-Learning}
\label{sec:olstar-tables}


We briefly explain the basics of the \lstar{} algorithm by~\citet{Angluin1987}.
The aim of \lstar{} is to construct a Mealy machine $M$ from a given function $L \colon X^* \to Y^*$ such that $\sem{M} = L$.
The function $L$ is ``closed-box'' and can only be interacted with as an oracle, meaning that $L(w)$ can be queried for each individual input word $w$.
Additionally, the \lstar{} algorithm uses an equivalence oracle to check whether an hypothesised machine $M$ is correct.
If the hypothesis is incorrect, i.e., $\sem{M} \neq L$, \lstar{} obtains a counterexample $w \in X^*$.
The following definitions are similar to the original paper on \lstar{} but specialised to Mealy machines.

\begin{definition}[Observation table]\label{def:table}
    An \emph{observation table (for $L$)} is a tuple $(S, E, T)$ where $S \subseteq X^*$ is a finite prefix-closed set of prefixes, $E \subseteq X^+$ is a finite set of suffixes, and $T$ is a function $T \colon S\,\cup\,S\!\cdot\!X \to (Y^*)^E$ such that $T(s)(e) = L(se)$.
\end{definition}

\begin{definition}[Closed and consistent]
    An observation table $(S, E, T)$ is
    \begin{itemize}[nosep]
        \item \emph{closed} if for all $s a \in S \cdot X$ there is $s' \in S$ such that $T(s') = T(s a)$,
        \item \emph{consistent} if for all $s, s' \in S$ with $T(s) = T(s')$ we have $T(s a) = T(s' a)$ for all $a \in X$.
    \end{itemize}
\end{definition}

\section{Output Decomposed Learning}
\label{sec:decompose}

Before describing our \olstar{} algorithm, we adopt the above definitions of the observation table and its properties to our setting of projections.

\begin{definition}[Projections on tables]\label{def:tableprojection}
    Given an observation table $OT = (S, E, T)$ and an output $y \in Y$, we define the projection of the table onto $y$ by $\oneact{y} OT = (S, E, \oneact{y} T)$ where $(\oneact{y} T)(s)(e) = \one{y}(T(s)(e))$.
\end{definition}

\begin{definition}[Output-closed and Output-consistent]
    An observation table $OT$ is
    \begin{itemize}[nosep]
        \item \emph{output-closed} if $\oneact{y} OT$ is closed for every $y \in Y$,
        \item \emph{output-consistent} if $\oneact{y} OT$ is consistent for every $y \in Y$.
    \end{itemize}
\end{definition}

These new notions are related to the regular closedness and consistency properties for observation tables in the following way.

\begin{lemma}
    Let $OT$ be an observation table, then:
    \begin{enumerate}[nosep]
        \item $OT$ is closed $\implies$ $OT$ is output-closed,
        \item $OT$ is consistent $\impliedby$ $OT$ is output-consistent.
    \end{enumerate}
\end{lemma}

An output-closed and output-consistent observation table defines a unique Mealy machine for every output. However, these machines may be in conflict with each other. There may be an input for which all components give the output 0, or there may be an input for which multiple components give the output 1. Therefore, we require that the components are consistent with each other.

\begin{definition}[Component-consistent]
    A family of Mealy machines $\{M^y\}_{y \in Y}$ with outputs in $\{0, 1\}$ is \emph{component consistent} if for every input word $w \in X^+$ there is a exactly one output $y \in Y$ such that $\last(\sem{M^y}(w)) = 1$.
\end{definition}

\begin{definition}
    Given a component-consistent family of Mealy machines $\{M^y\}_{y \in Y}$ with outputs in $\{0, 1\}$, the unique Mealy machine they represent is $M = \Pi_{y \in Y} (M^y, \one{y})$.
\end{definition}

\subsection{The \olstar{} Algorithm}

\algorithmref{alg:olstar} describes the \olstar{} algorithm in detail.
It mimics \lstar{} but with closedness and consistency replaced by output-closedness and output-consistency.
Additionally it has to check component-consistency.
The equivalence queries may only ask whether the (composed) hypothesis $H$ is equivalent to $M$; the teacher does not check the individual components $H^y$.

\begin{algorithm2e}[tb]
\caption{\olstar{}}
\label{alg:olstar}
\small 
\KwIn{Input alphabet $X$, MAT providing MQs and EQs}
\KwOut{Mealy machine equivalent to the target M}
$S \leftarrow \{\emptyword{}\}$;
$E \leftarrow X$;
Initialize $T$ with MQs, add observed outputs to $Y$\;
\Repeat{EQ(H) = YES}{
    \While{(S, E, T) is not output-closed or output-consistent}{
        \If{(S, E, T) is not output-closed}{
        find $y \in Y$, $s a \in S \cdot X$ such that there is no $s \in S$ with $\oneact{y} T(s a) = \oneact{y} T(s)$\;
        add $s a$ to $S$\;
        fill new rows of $T$ with MQs, add newly observed outputs to $Y$\;
        }
        \If{(S, E, T) is not output-consistent}{
            find $y \in Y$, $s, s' \in S$, $a \in X$, $e \in E$ such that $\oneact{y} T(s) = \oneact{y} T(s')$ but $\oneact{y} T(s a)(e) \neq \oneact{y} T(s' a)(e)$\;
            add $a e$ to $E$\;
            fill new columns of $T$ with MQs, add newly observed outputs to $Y$\;
            }
    }
    create family of Mealy machines $\{H^y\}_{y \in Y}$ based on $(S, E, T)$\;
    \uIf{$\{H^y\}_{y \in Y}$ is not component-consistent}{
        find input $w$ for which zero or multiple components output 1\;
        add suffixes of $w$ to $E$ until the defect is fixed\;
        fill new column with MQs, add newly observed outputs to $Y$\;
    }\Else{
        create Mealy machine $H = \Pi_{y \in Y} (H^y, \one{y})$\;
        \If{EQ(H) = w}{
            add suffixes of $w$ to $E$ until the defect is fixed\;
            fill new column with MQs, add newly observed outputs to $Y$\;
        }
    }
}
return $H$\;
\end{algorithm2e}

We do not require that the set of outputs $Y$ is known beforehand. We can observe the outputs that are added to the table, and use those as our output alphabet. Since our hypothesis can never return an output that we have not seen previously, the teacher can never return $\mathit{YES}$ for an equivalence query until we have seen all possible outputs of $M$.

There are some optimisations that can be done to reduce the number of queries we need to ask to learn $M$. We will mention two below.

Since there may be multiple output-closedness or output-consistency defects at once, our goal should be to resolve all of them, instead of solving the first defect we find. Trying to do this in the fewest number of tries reduces to the hitting set problem, which is NP-complete~\citep{Karp72}. Therefore, we choose the word that fixes the most defects in one go. This is a known approximation to the hitting set problem which works well in practice~\citep{Grossman1997}.

It can happen that our table is not only output-closed, but is also regularly closed. In this case, we immediately create the hypothesis and ask an equivalence query. This skips the checks for output-consistency and component-consistency. It should be noted that the table is always regularly consistent, since we never add a row to $S$ that is equivalent to one that is already in $S$.

\section{Empirical Evaluation}
\label{sec:eval}

In this section we present preliminary experiments we carried out to evaluate how well \olstar{} works. The experiments were run using our implementation\footnote{The code is available at \url{https://github.com/SCRK16/OLStar}.} of the algorithm using the LearnLib framework~\citep{IsbernerHS15}.
We use two sets of benchmarks.
The first are a set of Mealy machines where each machine consists of multiple, randomly generated, smaller machines sharing an input alphabet $\{a, b\}$, but all with unique outputs. Special inputs $L$ and $R$ allow for switching between which of the smaller machines is active. We expect \olstar{} to do well on this set of benchmarks, since the machines can be split up into their smaller components. For example, one of the machines has 10,000 states, but the projections only have a combined 95 states.

The second set of benchmarks are the ones designed by \citet{LabbafGHM23}. These benchmarks consist of Mealy machines that are the parallel interleaving of multiple smaller automata. The components have between 2 and 13 states. The interleavings then have between 4 and 4860 states. The interleavings can be decomposed into their components by looking at their inputs. It should be noted that many of the components share outputs. As such, we expect that \olstar{} may not always be able to find an effective decomposition based on outputs.

To measure how well the \olstar{} algorithm does, we ran it on these benchmarks and compared it to \lstar{}. For the equivalence queries we used randomised testing based on the \textsf{Wp}-method by~\citet{FujiwaraBKAG91} as implemented in LearnLib. We counted the number of membership queries and testing queries needed to learn the automata, excluding the last equivalence query. We also counted the number of input symbols needed for both the membership queries and testing queries. Both the learning and testing queries had their own cache, so repeated queries are only counted once. We compare the number of symbols required in \figureref{fig:comparison}. More statistics can be found in \appendixref{apd:figures}.

\subsection{Results}

\figureref{fig:comparison} shows the number of symbols required to learn a model for each instance.
The dashed plot shows the line where \lstar{} and \olstar{} pose the same number of queries, for every instance below that line we see that \olstar{} performs better.

\begin{figure}[htbp]
\floatconts
  {fig:comparison}
  {\caption{The number of symbols used by \lstar{} versus \olstar{}.}}
  {%
    \subfigure[Artificial benchmark]{\label{fig:artificialsymbols}%
      \includegraphics[width=0.43\linewidth]{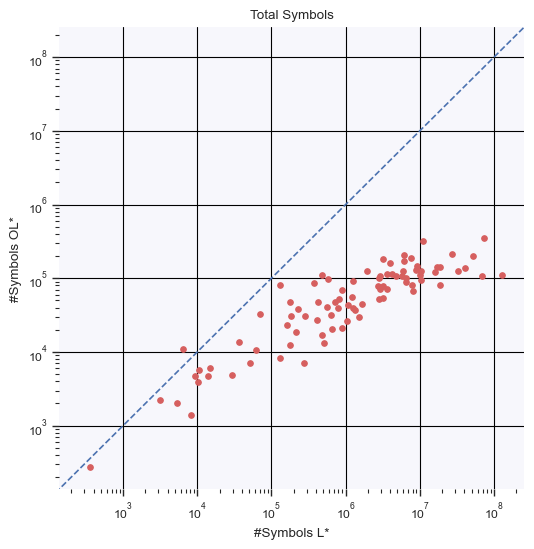}}\vspace{-6pt}%
    \qquad
    \subfigure[Labbaf et al. benchmark]{\label{fig:labbafsymbols}%
      \includegraphics[width=0.43\linewidth]{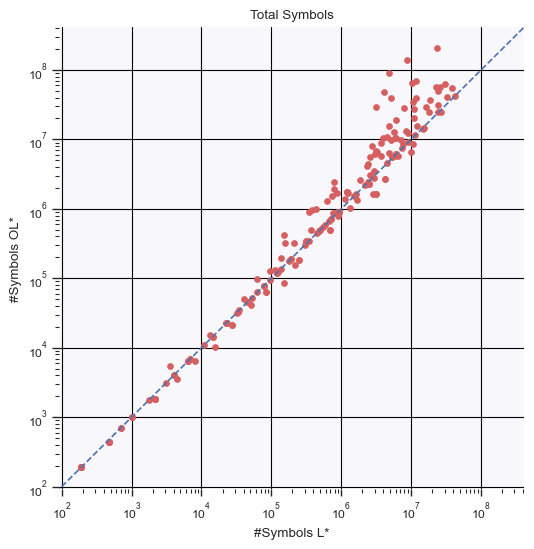}}\vspace{-6pt}
  }
\end{figure}

In the benchmark of~\citet{LabbafGHM23}, \olstar{} does not perform better than \lstar{}, this is to be expected as their components share the same outputs.
The reason \olstar{} does worse (as opposed to equal to \lstar{}) in these cases is because of the number of columns needed to make the table output-consistent. This is especially a problem when the target cannot be split up into smaller components, but the table is initially not closed in the regular sense. In these cases, \olstar{} does a lot of queries that ultimately lead to no benefit. When the target can be split up into smaller components, the extra columns needed for output-consistency are often compensated by the smaller number of rows needed to make the table output-closed.

\section{Conclusion and Discussion}

We have presented the algorithm \olstar{}, which learns a Mealy machine using a decomposition based on outputs in the Minimally Adequate Teacher framework. Although \olstar{} works well in some instances, it is still a work in progress and many improvements are conceivable.

The version of \olstar{} we have presented uses $\{\one{y} \colon Y \to \{0, 1\}\}_{y \in Y}$ as the set of jointly injective functions.
The generality of \definitionref{def:recomposition} and \lemmaref{lemma:reconstruction} hint at the possibility of using other decompositions in \olstar{}. In fact, \lstar{} itself is an instance of this: it uses a set with a single projection $\{\mathsf{id} \colon Y \to Y\}$.
Similarly, the work by~\citet{Moerman18} is an instance where a good set of jointly injective functions is already known to the learner.
A possible direction is to find the best possible set of jointly injective functions on-the-fly, although it seems that this is a computationally hard problem.

Another possibility is to combine decomposition based on outputs with other forms of decomposition to reduce the size of components even further. The approach of \citet{LabbafGHM23} is dual to our approach, decomposing the machine based on its inputs instead of its outputs.
We believe that these types of decomposition are orthogonal and could be used together in a learning algorithm.

Finally, we could adjust the Minimally Adequate Teacher framework to allow equivalence queries for individual components. Since the components themselves are smaller, these equivalence queries may be more efficient than equivalence queries for the whole system.


\small 
\bibliography{references}

\begin{thebibliography}{13}
\providecommand{\natexlab}[1]{#1}
\providecommand{\url}[1]{\texttt{#1}}
\expandafter\ifx\csname urlstyle\endcsname\relax
  \providecommand{\doi}[1]{doi: #1}\else
  \providecommand{\doi}{doi: \begingroup \urlstyle{rm}\Url}\fi

\bibitem[Angluin(1987)]{Angluin1987}
Dana Angluin.
\newblock Learning regular sets from queries and counterexamples.
\newblock \emph{Inf. Comput.}, 75\penalty0 (2):\penalty0 87--106, 1987.
\newblock URL \url{https://doi.org/10.1016/0890-5401(87)90052-6}.

\bibitem[Fiterau{-}Brostean et~al.(2023)Fiterau{-}Brostean, Jonsson, Sagonas, and T{\aa}quist]{Fiterau-Brostean23}
Paul Fiterau{-}Brostean, Bengt Jonsson, Konstantinos Sagonas, and Fredrik T{\aa}quist.
\newblock Automata-based automated detection of state machine bugs in protocol implementations.
\newblock In \emph{{NDSS}}. The Internet Society, 2023.
\newblock URL \url{https://www.ndss-symposium.org/ndss-paper/automata-based-automated-detection-of-state-machine-bugs-in-protocol-implementations/}.

\bibitem[Frohme and Steffen(2021)]{FrohmeS21}
Markus Frohme and Bernhard Steffen.
\newblock Compositional learning of mutually recursive procedural systems.
\newblock \emph{Int. J. Softw. Tools Technol. Transf.}, 23\penalty0 (4):\penalty0 521--543, 2021.
\newblock URL \url{https://doi.org/10.1007/s10009-021-00634-y}.

\bibitem[Fujiwara et~al.(1991)Fujiwara, von Bochmann, Khendek, Amalou, and Ghedamsi]{FujiwaraBKAG91}
Susumu Fujiwara, Gregor von Bochmann, Ferhat Khendek, Mokhtar Amalou, and Abderrazak Ghedamsi.
\newblock Test selection based on finite state models.
\newblock \emph{{IEEE} Trans. Software Eng.}, 17\penalty0 (6):\penalty0 591--603, 1991.
\newblock URL \url{https://doi.org/10.1109/32.87284}.

\bibitem[Grossman and Wool(1997)]{Grossman1997}
Tal Grossman and Avishai Wool.
\newblock Computational experience with approximation algorithms for the set covering problem.
\newblock \emph{European Journal of Operational Research}, 101\penalty0 (1):\penalty0 81--92, 1997.
\newblock ISSN 0377-2217.
\newblock URL \url{https://www.sciencedirect.com/science/article/pii/S0377221796001610}.

\bibitem[Isberner et~al.(2014)Isberner, Howar, and Steffen]{IsbernerHS14}
Malte Isberner, Falk Howar, and Bernhard Steffen.
\newblock The {TTT} algorithm: {A} redundancy-free approach to active automata learning.
\newblock In \emph{{RV}}, volume 8734 of \emph{Lecture Notes in Computer Science}, pages 307--322. Springer, 2014.
\newblock URL \url{https://doi.org/10.1007/978-3-319-11164-3\_26}.

\bibitem[Isberner et~al.(2015)Isberner, Howar, and Steffen]{IsbernerHS15}
Malte Isberner, Falk Howar, and Bernhard Steffen.
\newblock The open-source learnlib - {A} framework for active automata learning.
\newblock In \emph{{CAV} {(1)}}, volume 9206 of \emph{Lecture Notes in Computer Science}, pages 487--495. Springer, 2015.
\newblock URL \url{https://doi.org/10.1007/978-3-319-21690-4\_32}.

\bibitem[Karp(1972)]{Karp72}
Richard~M. Karp.
\newblock Reducibility among combinatorial problems.
\newblock In \emph{Complexity of Computer Computations}, The {IBM} Research Symposia Series, pages 85--103. Plenum Press, New York, 1972.
\newblock URL \url{https://doi.org/10.1007/978-1-4684-2001-2\_9}.

\bibitem[Labbaf et~al.(2023)Labbaf, Groote, Hojjat, and Mousavi]{LabbafGHM23}
Faezeh Labbaf, Jan~Friso Groote, Hossein Hojjat, and Mohammad~Reza Mousavi.
\newblock Compositional learning for interleaving parallel automata.
\newblock In \emph{FoSSaCS}, volume 13992 of \emph{Lecture Notes in Computer Science}, pages 413--435. Springer, 2023.
\newblock URL \url{https://doi.org/10.1007/978-3-031-30829-1\_20}.

\bibitem[Moerman(2018)]{Moerman18}
Joshua Moerman.
\newblock Learning product automata.
\newblock In \emph{{ICGI}}, volume~93 of \emph{Proceedings of Machine Learning Research}, pages 54--66. {PMLR}, 2018.
\newblock URL \url{http://proceedings.mlr.press/v93/moerman19a.html}.

\bibitem[Neele and Sammartino(2023)]{NeeleS23}
Thomas Neele and Matteo Sammartino.
\newblock Compositional automata learning of synchronous systems.
\newblock In \emph{{FASE}}, volume 13991 of \emph{Lecture Notes in Computer Science}, pages 47--66. Springer, 2023.
\newblock URL \url{https://doi.org/10.1007/978-3-031-30826-0\_3}.

\bibitem[Sanchez et~al.(2019)Sanchez, Groote, and Schiffelers]{SanchezGS19}
Lisette Sanchez, Jan~Friso Groote, and Ramon R.~H. Schiffelers.
\newblock Active learning of industrial software with data.
\newblock In \emph{{FSEN}}, volume 11761 of \emph{Lecture Notes in Computer Science}, pages 95--110. Springer, 2019.
\newblock URL \url{https://doi.org/10.1007/978-3-030-31517-7\_7}.

\bibitem[Vaandrager(2017)]{Vaandrager17}
Frits~W. Vaandrager.
\newblock Model learning.
\newblock \emph{Commun. {ACM}}, 60\penalty0 (2):\penalty0 86--95, 2017.
\newblock URL \url{https://doi.org/10.1145/2967606}.

\end{thebibliography}

\newpage 
\appendix
\section{Appendix}\label{apd:appendix}

\subsection{Extra figures}\label{apd:figures}

\begin{figure}[!htb]
\floatconts
  {fig:apd1}
  {\caption{Number of Membership + Testing queries (not counting individual symbols).}}
  {%
    \subfigure[Artificial benchmark]{\includegraphics[width=0.48\linewidth]{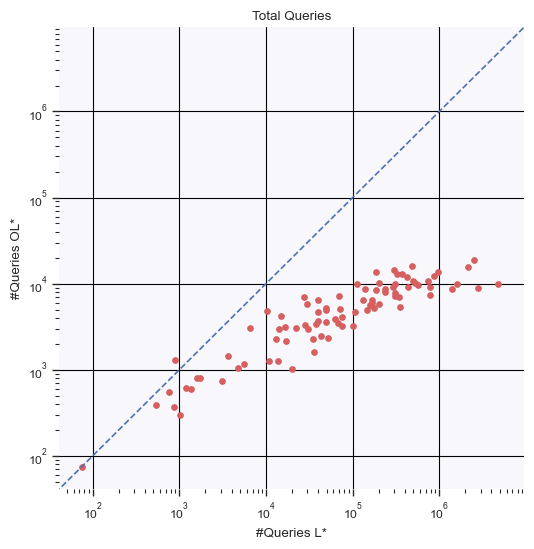}}%
    \hfill
    \subfigure[Labbaf et al. benchmark]{\includegraphics[width=0.48\linewidth]{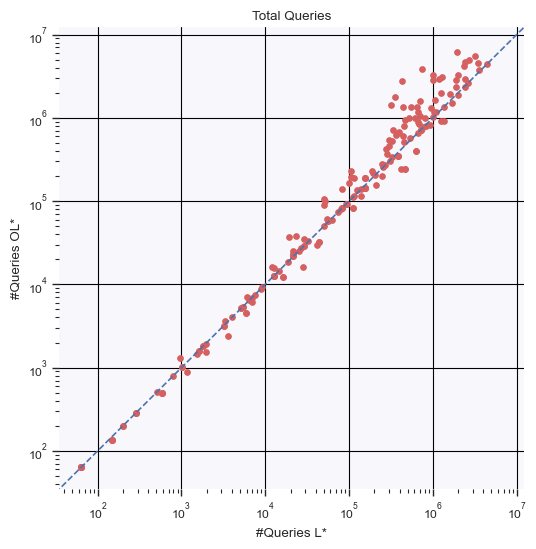}}\vspace{-12pt}
  }
\end{figure}

\begin{figure}[!htb]
\floatconts
  {fig:apd2}
  {\caption{Number of Membership queries (not counting individual symbols).}}
  {%
    \subfigure[Artificial benchmark]{\includegraphics[width=0.48\linewidth]{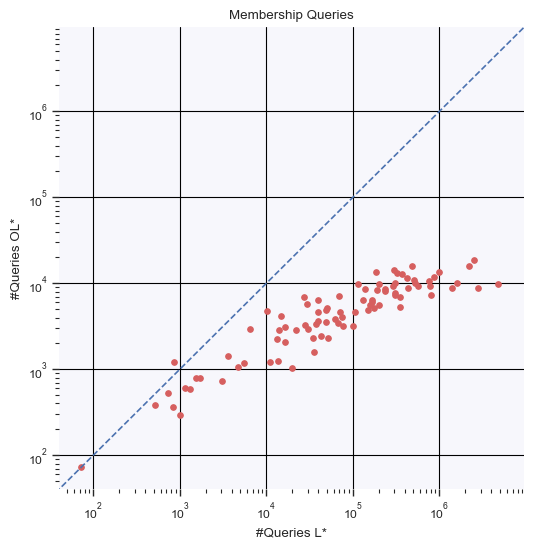}}%
    \hfill
    \subfigure[Labbaf et al. benchmark]{\includegraphics[width=0.48\linewidth]{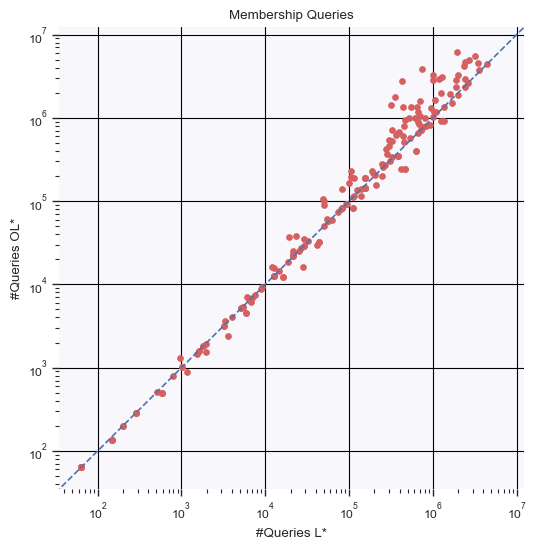}}\vspace{-12pt}
  }
\end{figure}

\begin{figure}[!htb]
\floatconts
  {fig:apd3}
  {\caption{Number of Testing queries (not counting individual symbols).}}
  {%
    \subfigure[Artificial benchmark]{\includegraphics[width=0.48\linewidth]{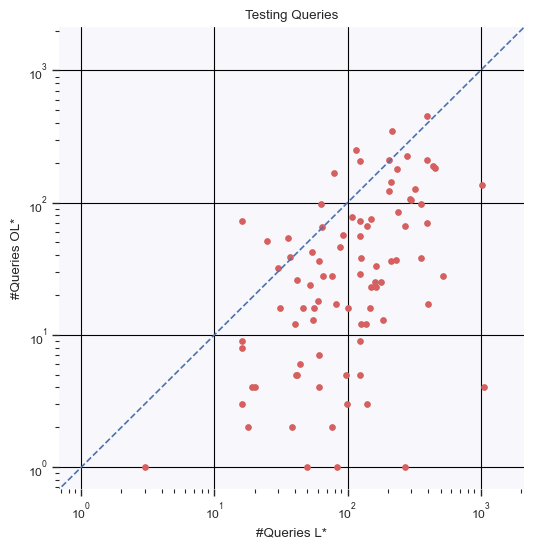}}%
    \hfill
    \subfigure[Labbaf et al. benchmark (Note: There are very few machines in the benchmark for which testing queries are needed)]{\includegraphics[width=0.48\linewidth]{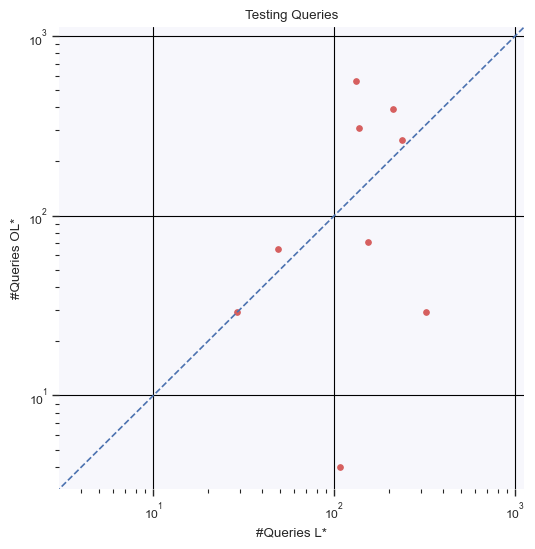}}\vspace{-12pt}
  }
\end{figure}

\begin{figure}[!htb]
\floatconts
  {fig:apd4}
  {\caption{Number of Equivalence queries.}}
  {%
    \subfigure[Artificial benchmark]{\includegraphics[width=0.48\linewidth]{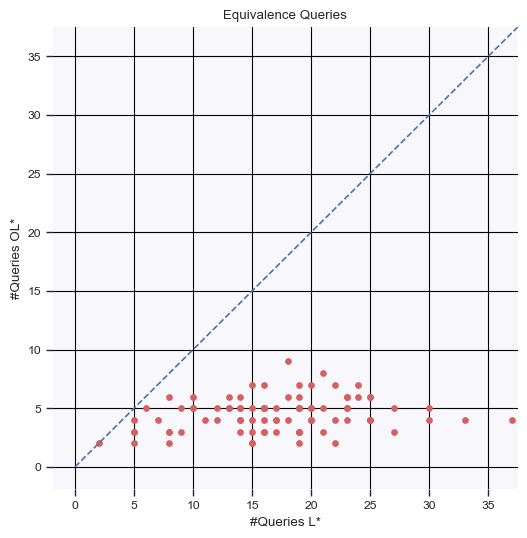}}%
    \hfill
    \subfigure[Labbaf et al. benchmark]{\includegraphics[width=0.48\linewidth]{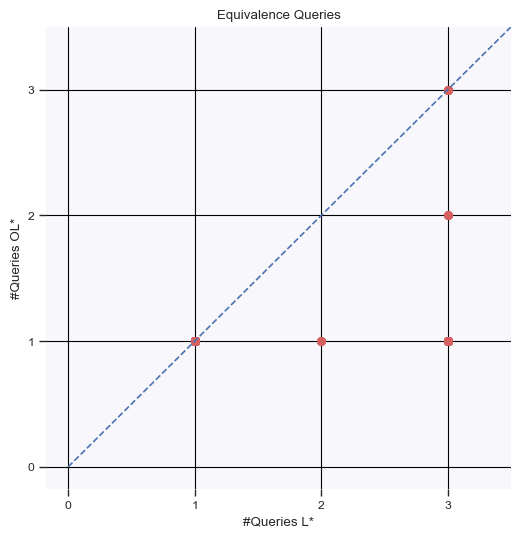}}\vspace{-12pt}
  }
\end{figure}

%
%

\end{document}